# A Radio Based Intelligent Railway Grade Crossing System to Avoid Collision

Sheikh Shanawaz Mostafa [1], Md. Mahbub Hossian [2], Khondker Jahid Reza [3], Gazi Maniur Rashid [4]

[1] Electronics and Communication Engineering Discipline, Khulna University,
Khulna-9208, Bangladesh.

[2] Electronics and Communication Engineering Discipline, Khulna University,
Khulna-9208, Bangladesh.

[3] Electronics and Communication Engineering Discipline, Khulna University,
Khulna-9208, Bangladesh.

[4] Electronics and Communication Engineering Discipline, Khulna University,
Khulna-9208, Bangladesh.

**Abstract**
Railway grade crossing is become the major headache for the transportation system. This paper describes an intelligent railway crossing control system for multiple tracks that features a controller which receives messages from incoming and outgoing trains by sensors. These messages contain detail information including the direction and identity of a train. Depending on those messages the controller device decides whenever the railroad crossing gate will close or open.

## 1. Introduction

A railway grade crossing is a point at which a railway and a road intersect on the same level [4]. In many situations two or more tracks may cross a highway. It creates problem for a bus or truck or other vehicles to clear both tracks. In recent five years, more than three hundred people are killed and another about two hundred are injured due to railway grade crossing accidents [5]. Whether accidents are caused by the negligence of the men, undesirable weather conditions, inadequate traffic planning. A large number of grade crossings exist without any gate. There is an inherent unreliability in the present manual system. Other hand, 'constant warning time' system provides fixed warning time regardless of an approaching train. So that, for a faster train the safety devices are activated earlier and for a slower train activated later. Installing and maintaining costs are too much for these above two systems.

Global Positioning System (GPS) based rail crossing system is discussed by many authors. Some author discusses the use of data obtained from GPS devices located on trains or at railroad crossings to provide train's approaching information [2]. Other installed GPS receiver is on top of a train and used to obtain information concerning the train's speed and position [3]. And a radio link based system describe in by [1] .But all the systems are not useful for deriving arrival time and train speed information for multiple trains at a time. And they cannot derive information concerning the identity and status of individual trains (like entering or leaving the system).

According to an aspect of the present situation, this paper describes an intelligent railroad crossing system. That can be used to avoid collision by road vehicles approaching towards multiple tracks railroad crossing.

## 2. Design Pattern

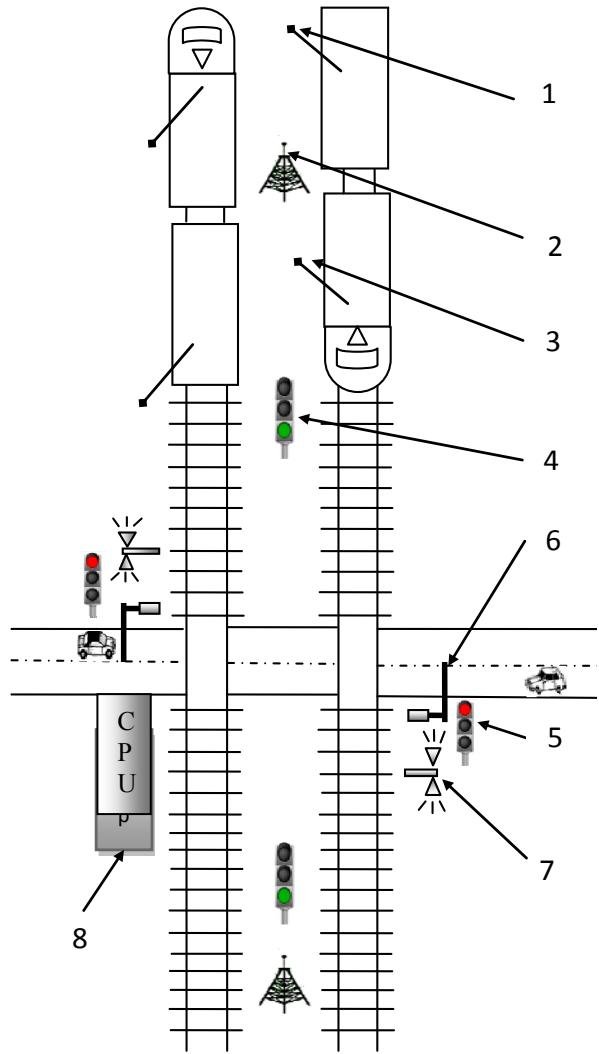

Fig. 1 Top view of the intelligent grade crossing system

This automatic railway crossing gate uses radio link for identification, information of approaching and outgoing trains. The train has two transmitters at the beginning (3 in Fig. 1) and end (1in Fig. 1) which transmit an identical packet that can be identified by the sensor (2 in Fig. 1). This packet transmitted through a radio link and received by sensor. Then the sensor sends the information of the packet to cpu (8 in Fig. 1) where the controlling procedure is processed. This cpu consists of several thing like packet identification signalling and gate controller device. After receiving the packet the cpu changes the signal & gate status from the packet type & algorithm stored in the cpu. This system consists of two kinds of signalling posts for safety purpose; one for train (4 in Fig. 1) and another for street traffic (5 in Fig. 1). There is also an audio signalling system (7 in Fig. 1) for unconscious users. The system consists of half barrier gates for escaping trapped vehicles. When there is a train in the system it shuts the gate and opens it after leaving the system.

## 3. System Block

3.1 Transmitter

Two transmitters (1 & 3 in Fig. 1) are mounted in

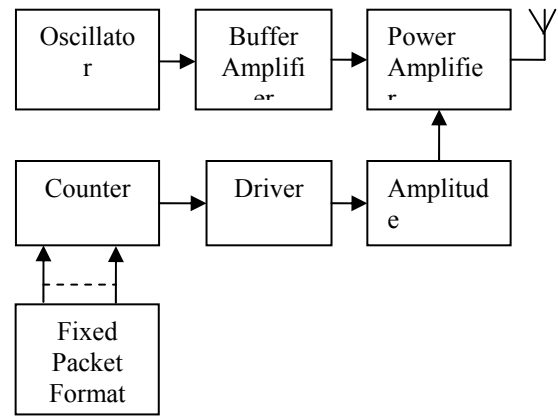

Fig. 2 Transmitter

every train. One is in front and another is back of the train. Each train has a fixed identification number. The train transmits a packet consisting of this identification number along with the position of the transmitter (front or back). The block diagram of the transmitter is shown in Fig. 2.

3.2 Sensor

Sensors (2 in Fig. 1) are situated besides the two railway tracks. Normally, two sensors are placed both side of the railway crossing gate. Sensors receive the packets those are sent by the train's transmitters. Sensors are one kind of radio antenna.

## 3.3 Signal

Two types of visual signals (4 and 5 in Fig. 1) are used for safety purposes. One is street signal for road traffic and another is train itself. An audio signal is also introduced in the system.

## 3.4 Gate

The gates (6 in Fig. 1) are consists of mechanical and electrical subsystems. Two half barricade gates are the mechanical portions of the gate. Gates also have electrical portion equipped with driven motor controlled by the cpu.

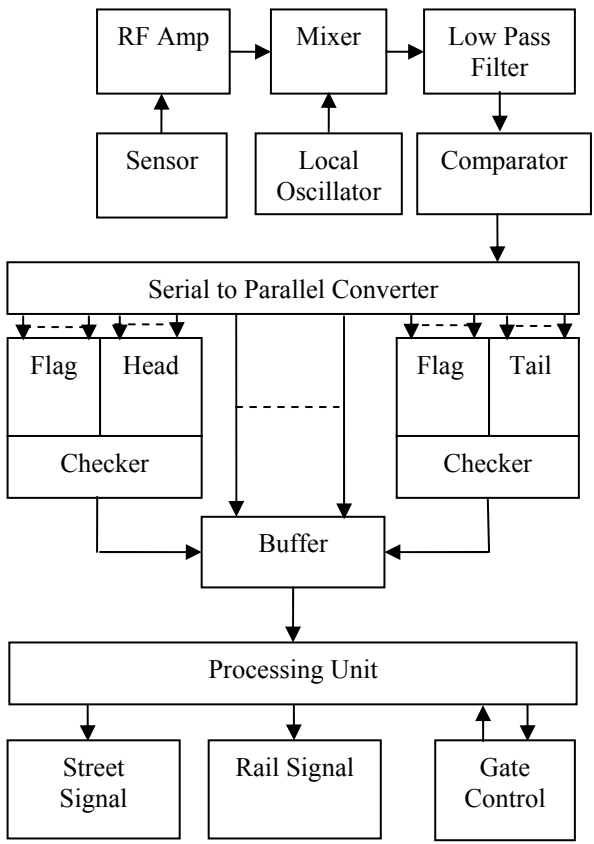

Fig 3 Central processing unit

## 3.5 Central Processing Unit (cpu)

Central processing unit (8 in Fig. 1) or central controlling unit connects with three other systems: sensor, gate and signaling. Signal coming from the sensor is processed and decision is sent to the gate and signal.

First portion of this Central processing unit is a radio receiver consists of Radio Frequency Amplifier (RF Amp), Mixer, Low Pass Filter, Sensor, Local Oscillator, Comparator. Then the second part Serial to Parallel Converter with a buffering system built within it. And the processing unit is actually a computer. We used parallel port to input and output data into computer. Visual basic is used for programming the system. The program first take input from parallel port. Then execute the program according working flow of the system. And send out put.

## 4. Working Flow Diagram

The processing unit takes decision according to its own algorithm. The flow chart of decision making algorithm is shown in Fig. 4 where sg_st and sg_tr mention street and train signal status respectively. Both street signal and train signal have two situations {g,r} where g stands for green signal and r for red signal. Here, g_s is the gate status and g_s $\in$ {o,c} for closing the gate it is c and for opening o. Variable x is a three dimensional array which is actually memory of the system. On the other hand i, j, k are incremental variable and i $\in$ {1, 2, ..., m}, where m is total number of memory. The packet has two information; train id (tr_id) and phase (d) of the train. Here, d $\in$ {h,t} it is h if the packet generate from head of the train otherwise it is t. Variable s indicates at which sensor the signal is acquired from. Variable g_sr $\in$ {o,c} o when the gate is really open and c when the gate is really close it is actually a feedback status from gate.

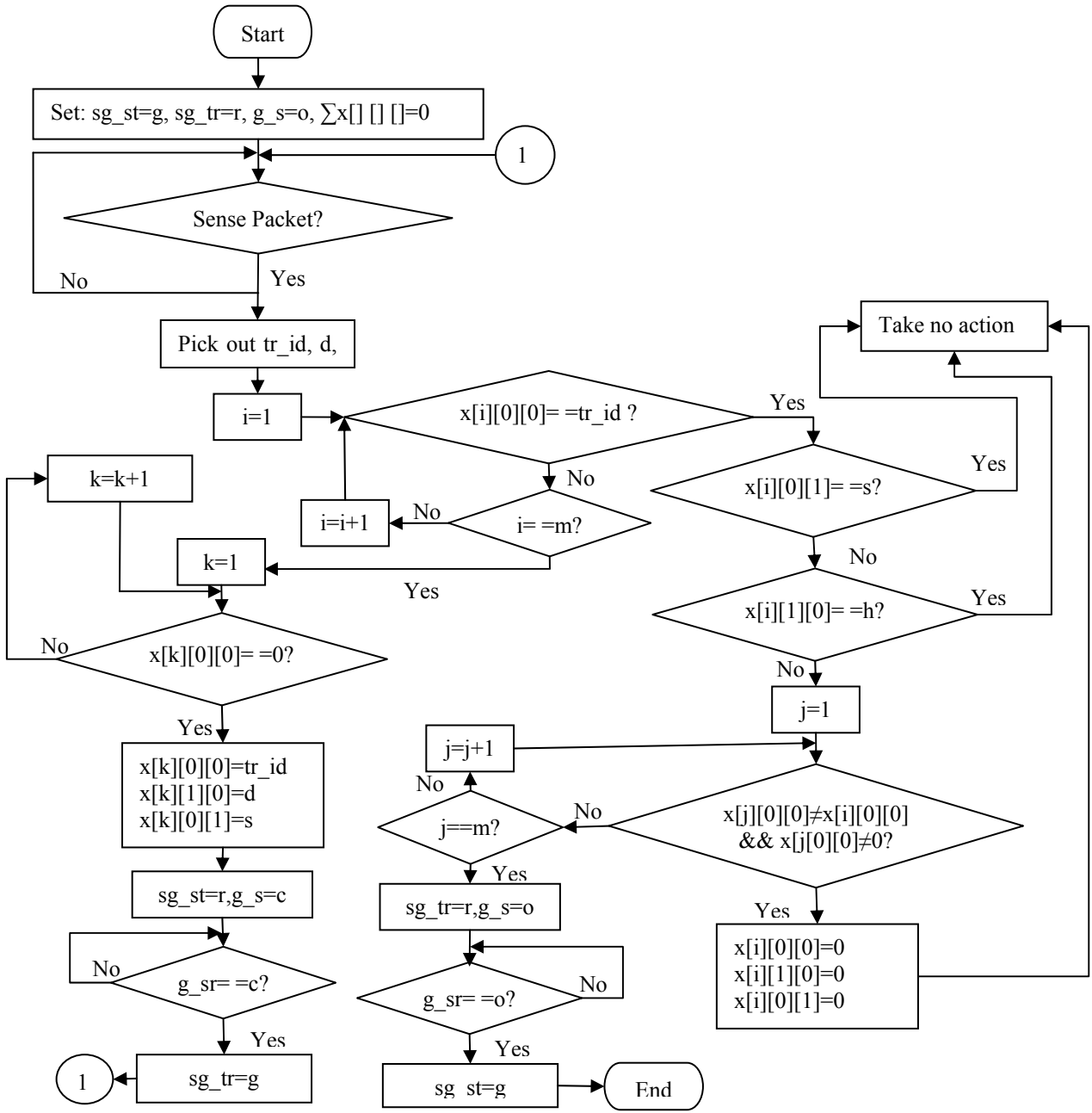

Fig. 4 Flow chart of the decision making of the system

## 5. Conclusion

An automatic railway crossing gate control system using radio link has been proposed in this paper for reducing the accidents. This system can be implemented both on single and multiple tracks. The main facility of this system is that it can be merged with the existing system. The initial cost of the system is a little high but maintenance cost is very low. Power consumption of the system is low. Using solar energy for providing the power can make the system more fruitful to the rural areas. In future the system would be web based. This system is standalone system and can work 24×7 which is impossible for any man operated system.

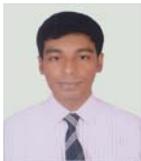
**Sheikh Shanawaz Mostafa** completed B.Sc. in Electronics and Communication Engineering Discipline in Khulna University-9208, Khulna, Bangladesh. His current research interest is wireless communication, modulation and biomedical signal processing. His numbers of published papers are four, among them international recognized journal and proceedings of local conference.

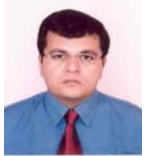
**Md. Mahbub Hossain** completed his B.Sc Engineering degree in Electronics and Communication in the year of 2003 from Khulna University, Khulna-9208, Bangladesh. He is now the faculty member of Electronics and Communication Engineering Discipline, Khulna University, Khulna-9208, Bangladesh. His current research interest is wireless communication, modulation techniques, channel coding and fading. His number of published papers are 11 among them international recognized journal and proceedings of international and local conference.

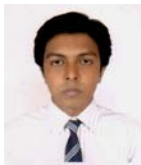
**Khondker Jahid Reza** is serving as a Engineer. He completed his B.Sc. in Electronics and Communication Engineering Discipline in Khulna University, Khulna, Bangladesh. His current research interest is wireless communication, modulation and sensor networks.

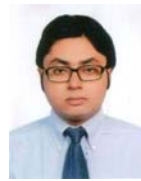
**Gazi Maniur Rashid** is currently pursuing B.Sc. in Electronics and Communication Engineering Discipline in Khulna University, Khulna, Bangladesh. His current research interest is wireless communication, modulation.